\documentstyle[preprint,aps,epsf]{revtex}

\begin{document}

\title{
Heavy Hybrids with Constituent Gluons}

\author{Eric S. Swanson$^{1,2}$ and Adam P. Szczepaniak$^{3}$}

\address{$^1$\ 
   Department of Physics,
   North Carolina State University,
   Raleigh, North Carolina 27695-8202\\
         $^2$\ 
   Jefferson Laboratory, 
   12000 Jefferson Avenue, 
   Newport News, VA 23606\\
         $^3$\ 
   Physics Department,
   Indiana University,
   Bloomington, Indiana   47405-4202\\
   } 

\maketitle

\begin{abstract}
Hybrid meson energies are calculated in the static quark limit with the
Dynamical Quark Model (DQM). In the DQM, transverse gluons are 
represented as effective constituents with a dynamically generated mass.
Hybrid masses are
determined within the Tamm-Dancoff approximation for the resulting
relativistic Salpeter equation. Although the general features of the 
adiabatic potential surfaces correspond with lattice data, the results
disagree on level orderings. Similar problems appear to exist in all 
constituent glue models of hybrids. We conclude that constituent gluons
do not accurately represent soft gluonic degrees of freedom. The steps 
necessary to correct this deficiency are discussed.

\end{abstract}
\date{Dec, 1997}
\narrowtext

\section{Introduction} 

A decade of experimental signals\cite{hist} for QCD hybrids 
(in particular with $J^{PC} = 1^{-+}$) has culminated in the 
claimed observation of three such states
at Brookhaven\cite{e852} in the last year.
The question of the nature of QCD hybrids has thus become topical.
Furthermore, lattice gauge calculations are now at the point of accurately
determining light hybrid masses. In view of these developments, it is of
interest to compare models of strong (low energy) QCD  
with lattice data to determine their viability and to explicate
and guide current experimental efforts.

It is often stated that a hybrid is a hadron consisting of valence quarks
and glue. However, one must specify what is meant by the notion of valence 
glue for this statement to be useful.
There are two broad ideas in this regard: it is some sort of string or flux
tube\cite{strings,IP} or it is an effective constituent confined by a 
bag\cite{B,HHKR} or potential\cite{HM,B2,SC}. 

As an example of the importance of choosing correct degrees of freedom, we
mention the simple problem of determining the
number of components of a constituent gluon.
It has been suggested that a massive
constituent gluon should be transverse so as to maintain consistency with
Yang's theorem\cite{B2}. However it was noted that this is inconsistent
with the requirements of Lorentz invariance. 
Thus, for example, $J = 1$ glueballs are
expected to exist and lattice calculations indicate that they are quite heavy
(roughly 3 GeV)\cite{lattgb}.  Such a state may not be constructed from two
transverse constituent gluons (Yang's theorem) and therefore may be expected to
have a mass of roughly $3 m_g \sim 3$ GeV. However massive vector gluons have
no such constraint and one therefore expects them to have a mass of 
approximately $2 m_g \sim 2$ GeV.

The nature of the appropriate effective degrees of freedom for glue can only be
determined by a long process of calculation and comparison with experimental
and lattice data.
There are, however, a few indications that low energy glue is string-like. 
Perhaps the most compelling of these are lattice calculations of energy,
action, or field densities between static color sources which are
reminiscent of flux tubes\cite{fts}. An intriguing clue is also provided
by the spin splittings of heavy quarkonia. 
It is known that an effective interaction free of long range exchange
spin-orbit coupling is needed to reproduce the mass splitting of the $^3P_J$
 heavy quarkonium multiplets. However, an analysis of QCD in the heavy quark limit
convincingly demonstrates that obtaining such an effective potential
requires that low energy glue must be string-like\cite{ss3}.

Alternatively, pointlike models of low energy glue have a 
long history, 
originating with MIT bag model calculations of Barnes\cite{B} and 
others\cite{HHKR}.
Horn and Mandula\cite{HM} were the first to consider a potential constituent 
glue model
of hybrids. Their hybrids consisted of constituent quarks and pointlike,  
massless, spinless, and colorless glue in a nonrelativistic potential model.
The confining potential was taken to be linear with a string tension given
by the ratio of color Casimir operators: $b_{qg} = 9/8 b_{q\bar q}$. The
authors noted that the two body $q \bar q$ potential is anti-confining in the
color octet channel and has a repulsive Coulomb spike at short distances. 
They argued that this is
unphysical and hence choose to neglect this term in the interaction. It is 
clear that a great many simplifying assumptions have gone into the 
construction of this model. It is our purpose to compare a more 
sophisticated version of the model to lattice data to learn something about
these assumptions.

In the following we employ a model field theoretic Hamiltonian of QCD. The model
incorporates linear confinement at low energy and evolves into perturbative QCD
at high energy. A nontrivial vacuum is used to generate constituent quark 
and gluon masses. The eigenvalue equation is  derived for a $q\bar q g$ 
system where the quarks are static. The resulting adiabatic potential surfaces
are then compared to recent lattice results. We conclude that the simple picture
of glue as a pointlike constituent particle reproduces the general behaviour
of the lattice results but fails to yield the correct level orderings.
Furthermore, other models which regard the gluonic degrees of freedom as 
pointlike (eg., \cite{bcs}, \cite{HM}) do not contain sufficient degrees of
freedom to generate all of the adiabatic surfaces. Thus constituent glue 
models appear to fail to describe hybrids. This stands in contrast to 
string or bag-like models which, although disagreeing on details, capture 
the rough structure of the lattice data.

\section{A Constituent Glue Model of Static Hybrids}

\subsection{The Dynamical Quark Model}

The starting point for our description of hybrids is the following model
Hamiltonian:

\begin{eqnarray}
H &=& \int d{\bf x} \psi^{\dag}({\bf x})\left[ -i{\vec \alpha} \cdot
{\vec \nabla} + \beta m \right]
\psi({\bf x})  + {1\over 2} \int d{\bf x} \left[ |{\bf E}^A({\bf x})|^2
+  |{\bf B}^A(x) |^2  \right] \nonumber \\
 &+& {1\over 2}\int d{\bf x} d{\bf y} \rho^A({\bf x})V(|{\bf x} - 
{\bf y}|) \rho^A ({\bf y})
\end{eqnarray}
  
\noindent
where the color charge density is $\rho^A({\bf x}) = \psi^{\dag}({\bf x})
{\rm T}^A \psi({\bf x}) - f^{ABC} {\bf A}^B({\bf x})\cdot{\bf E}^C({\bf x})$ and  the
potential is given by

\begin{equation}
V(r) = {\alpha_s \over r} - {2 {N_c b} \over {N_c^2 - 1}} r \,
\left(1 - {\rm e}^{-\Lambda_{UV} r } \right)
\end{equation}

\noindent
and $N_c = 3$. The quark mass appearing in this Hamiltonian is the current 
mass. To be phenomenologically successful, constituent quark masses must
be generated in some way. This may be achieved by employing a BCS vacuum
Ansatz; the gap equation which follows from minimizing the vacuum energy 
density, $\langle\Omega\vert H \vert \Omega\rangle$, (where 
$\vert \Omega\rangle$ 
represents the BCS trial vacuum) gives rise to a low energy constituent 
quark mass
of roughly 200 MeV\cite{ss2}. A similar calculation in the glue sector 
yields a gluon dispersion relation which is well-approximated by

\begin{equation}
  \omega(k) = \sqrt{ k^2 + m_g^2 {\rm e}^{-k/\kappa}}
\end{equation}
\noindent

\noindent
with $m_g = 800$ MeV and $\kappa = 6.5$ GeV. One sees that a
constituent gluon mass of approximately 800 MeV has been generated\cite{ssjc}.
Hadrons are then constructed on top of the BCS vacuum $\vert \Omega\rangle$ 
by employing a basis truncation (typically Tamm Dancoff or Random Phase) and 
solving the resulting Bethe-Salpeter equation.  This approach has been used
to derive the low lying spectrum of glueballs and agrees remarkably well with
lattice data\cite{ssjc}. It should be noted that the dynamical gluons are
transverse so that Yang's theorem holds and the difficulty mentioned in the
Introduction does not arise.

In the following, the parameters of the model are fixed to the $q\bar q$ 
potential derived in the lattice calculation of Juge, Kuti, and 
Morningstar\cite{jkm} (see Fig. 3 below).  
The fit yields $\alpha_s = 0.29$ and $b = 0.24$ GeV$^2$.
The final parameter, $\Lambda_{UV}$ serves as an ultraviolet cutoff on 
the linear
confinement potential. Its value was determined in Ref. \cite{ssjc} by fitting
the gluon condensate, and will be set at 4 GeV in the following. We shall
consider the static quark limit in the remainder of this work so that quark
masses (and the quark sector of the BCS vacuum) become irrelevant.

The model presented here may be considered as a simplified version of the
Coulomb gauge QCD Hamiltonian where the effects of nonabelian gauge couplings 
have been modeled by the linear confinement term. Furthermore, second order
transverse gluon exchange is suppressed by the heavy quark masses. In principle,
this approach allows the elimination of the ultraviolet scale $\Lambda_{UV}$ via
renormalization. A method for achieving this which is appropriate for 
nonperturbative Hamiltonian-based calculations is described in Ref. \cite{ss2}.

\subsection{Static Hybrids}

In the Tamm Dancoff approximation, hybrids are constructed as $q \bar q g$
excitations of the BCS vacuum. For the heavy hybrids considered here, the 
(anti)quarks serve as static
color sources (sinks) and the gluons are constituent particles as described 
above.  We choose to work in the ``diatomic molecule" basis because this 
facilitates comparison with the lattice results of Ref. \cite{jkm}. Thus 
the hybrid state may be written as

\begin{equation}
|\vec R\, n_g; j_g\, \Lambda \,\xi\rangle = \int d \vec k 
\varphi_{n_g j_g}(k)
{\cal D}^{j_g}_{\mu\Lambda}(\hat R) {\cal D}^{j_g *}_{\mu \lambda'}(\hat k) 
\sqrt{2 j_g +1\over 4 \pi} \chi_{\lambda \lambda'}^\xi {\rm T^A_{ab} \over 2}
b^\dagger_{\vec R/2,a} d^\dagger_{-\vec R/2,b} a^\dagger_{k,A,\lambda} \vert \Omega 
\rangle. \label{basis}
\end{equation}

\noindent
Wigner rotation functions are written as ${\cal D}$ in this  expression, 
$R$ is the distance between the $q\bar q$ pair, $\lambda$ is the gluon
helicity, and $\varphi$ is the radial hybrid wavefunction in momentum
space. The gluon polarization wavefunction, $\chi^\xi_{\lambda\lambda'}$ is given by 
 $\chi^1_{\lambda\lambda'} = \delta_{\lambda\lambda'}/\sqrt{2}$ and 
$\chi^{-1}_{\lambda\lambda'} = \lambda \delta_{\lambda\lambda'}/\sqrt{2}$.
The two cases, $\xi =$1, -1, represent transverse magnetic and transverse 
electric hybrids respectively. Finally, $\Lambda$ is the projection of the
gluonic angular momentum onto the $q\bar q$ axis, $j_g$ is the total 
angular momentum of the gluon, and $n_g$ labels the radial basis state. We note that 
employing
helicity basis gluon creation operators makes this expression significantly
more compact than the canonical basis.

The Salpeter equation which follows from this ansatz and the Hamiltonian in
Eqn. (1) may be obtained from the following matrix element:

\begin{eqnarray}
\langle \vec R'\, n_g'; j_g'\, \Lambda'\, \xi'\vert &H& \vert \vec R\, 
n_g; j_g\, \Lambda\, \xi \rangle 
= \int d\vec k \,\varphi_{n_g' j_g'}^*(k) \varphi_{n_g j_g}(k) {1\over 2}[ \omega(k) + {k^2\over \omega(k)}]  \nonumber \\
&+& {3 \over 8} \int \int d \vec q d \vec k\, \varphi_{n_g' j_g'}^*(k) 
\varphi_{n_g j_g}(k)\, V(k-q) \left[ {\omega(k)^2 + \omega(q)^2 \over
\omega(k) \omega(q)} (1 + (\hat q \cdot \hat k)^2) \right]  \nonumber \\
&-& {3 \over 4} \int\int d \vec q d \vec k\, \varphi_{n_g' j_g'}^*(q)\varphi_{n_g j_g}(k)
\, V(k-q) \left({\rm e}^{i {R\over 2} \cdot (k-q)}
 + {\rm e}^{-i {R\over 2}\cdot (k-q)} \right) {\omega(k) + \omega(q)\over 
\sqrt{\omega(k) \omega(q)}}\times \nonumber \\
&&\times {\cal D}^{j_g*}_{\Lambda \lambda'}(\hat k) 
{\cal D}^{j_g'}_{\Lambda'\sigma'}(\hat q) \sqrt{2 j_g + 1 \over 4 \pi}
\sqrt{2 j_g'+1\over 4 \pi}
\chi_{\lambda \lambda'}^\xi \chi_{\sigma \sigma'}^{\xi'} {\cal D}^1_{\mu \lambda}(\hat k) {\cal D}^{1*}_{\mu\sigma}(\hat q)
\nonumber \\
&+& \left({1\over 6} V(R) + {4\over 3}V(0)\right) {\cal N}_{n_g' j_g' n_g j_g} \delta_{j_g' j_g} 
\delta_{\Lambda' \Lambda} \delta_{\xi' \xi} 
\end{eqnarray}

\noindent
An overall $\delta(\vec R' - \vec R)$ is understood and ${\cal N}_{n_g' j_g' n_g j_g} = \int dk k^2 
\varphi_{n_g' j_g'}^*(k) \varphi_{n_g j_g}(k)$ is the
wavefunction normalization factor. The two extra Wigner rotation matrices arise
from converting the Cartesian basis implicit in Eqn. (1) and in the expression
for the color current to the helicity basis, $a_{k i A} = \epsilon_H^i(k\lambda)
a_{k \lambda A} = {\cal D}^1_{m \lambda}(\phi,\theta, -\phi) \epsilon_C^i(m)
a_{k \lambda A}$, where $\epsilon_C$ and $\epsilon_H$ are canonical and 
helicity polarization vectors respectively.

The first term in this expression is the gluon kinetic energy, the second
is the gluon self energy, the third is the gluon potential, and the fourth
is the $q \bar q$ potential and self energies for static quarks in a color
octet. The presence of the gluon and quark self energies assures the 
infrared finiteness of the Salpeter equation. This appears to be a general 
feature of color singlet states in our approach\cite{ssjc}.

The last task is to identify the diatomic quantum numbers used to label the 
states.
In the Jacob-Wick convention the action of parity on glue is given by 

\begin{equation}
P a_{\vec k,\lambda,A} P^\dagger = \eta^P_g{\rm e}^{-2 i \lambda \phi} a_{-\vec k,
-\lambda,A}
\end{equation}

\noindent
where $\phi$ is the azimuthal angle of $\hat k$ and $\eta^P_g=-1$ is the intrinsic gluon parity. 
Thus the hybrid states given in Eqn. (4) are eigenstates of gluonic parity with 

\begin{equation}
P_g \vert \vec R n_g; j_g \Lambda \xi\rangle = \xi \eta^P_g (-)^{j_g+1}\vert \vec R n_g; 
j_g \Lambda \xi \rangle.
\end{equation}

\noindent
A reflection of the glue degrees of freedom through a plane containing the 
$q\bar q$ axis leaves
the Hamiltonian invariant and when acting on the states it takes $\Lambda \rightarrow -\Lambda$. 
For $\vert \Lambda \vert > 0$ one thus has doubly degenerate states. 
We call this operation  $Y$-parity  and perform it by setting
$\vec R \to R \hat z$ and taking $y \rightarrow -y$ which may be achieved
by a parity operation followed by a rotation through $\pi$ about the y-axis.
The action of $Y$ on a single transverse gluon state, 
$\vert k \lambda A\rangle = a^\dagger_{k \lambda A} \vert 0\rangle$ is 
therefore given by,

\begin{equation}
Y\vert k \lambda A\rangle = {\rm e}^{-i \pi J_y} P \vert k\lambda A\rangle =
 \eta^P_g{\rm e}^{2 i \lambda \phi}\vert k', -\lambda, A\rangle,
\end{equation}

\noindent
where $\vec k' = (k_x,-k_y,k_z)$. For the states defined in
 Eq.~(\ref{basis}) one has  

\begin{equation}
Y\vert \vec R n_g; j_g \Lambda \xi \rangle = \xi \eta^P_g (-)^{\Lambda+1} \vert \vec R 
n_g;j_g \, -\Lambda \xi \rangle,
\end{equation}
where  the relations ${\cal D}^j_{\mu -\mu'}(k') = 
(-)^{2j + \mu + \mu'} {\rm e}^{-2 i \mu' \phi} {\cal D}^j_{-\mu \mu'}(k)$
and ${\cal D}^j_{\mu\mu'}(\hat z) = \delta_{\mu\mu'}$ were used.

\noindent
For $\Lambda \ne 0$ the Y-diagonal states may thus be written as

\begin{equation}
\vert \vec R n_g; j_g |\Lambda| \xi; \eta_Y \rangle  = {1\over \sqrt{2}} \Bigg(
\vert \vec R n_g; j_g |\Lambda| \xi \rangle + \eta_Y \vert \vec R n_g; j_g 
-|\Lambda| \xi \rangle \Bigg), \label{Ybasis}
\end{equation}

\noindent
where $\eta_Y = \pm 1$ and, 

\begin{equation}
Y \vert \vec R n_g; j_g |\Lambda| \xi; \eta_Y \rangle  = \xi \eta^P_g \eta_Y (-)^{\Lambda +1}
  \vert \vec R n_g; j_g |\Lambda| \xi; \eta_Y \rangle.
\end{equation}
For $\Lambda=0$ we simply have,

\begin{equation}
Y \vert \vec R n_g; j_g 0\xi \rangle = 
 -\xi \eta^P_g \vert \vec R 
n_g;j_g 0 \xi \rangle.
\end{equation}

To allow for easy comparison with conventions used elsewhere, we summarize the  
transformation properties of $\bar q q g$  states under total parity,  
$P=P_q P_g$, and charge conjugation $C = C_q C_g$, including the quark degrees 
of freedom. 

\begin{equation}
P  \vert \vec R n_g; j_g |\Lambda| \xi; \eta_Y \rangle =   \xi \eta^P_g \eta^P_{\bar qq} (-)^{j_g+1}
\vert -\vec R n_g; j_g |\Lambda| \xi; \eta_Y \rangle
\end{equation}

\begin{equation}
C  \vert \vec R n_g; j_g |\Lambda| \xi; \eta_Y \rangle = \eta^C_g \eta^C_{\bar qq} 
(-)^{S_{\bar q q}+1}
  \vert -\vec R n_g; j_g |\Lambda| \xi; \eta_Y \rangle
\end{equation}
where $\eta^P_{\bar q q}$ = $\eta^C_{q\bar q}$ = $\eta^C_g$ = -1. The states 
 introduced in 
Eq.~(\ref{Ybasis}) are therefore also eigenstates of combined $PC$, with 

 \begin{equation}
P C  \vert \vec R n_g; j_g |\Lambda| \xi; \eta_Y \rangle = \xi \eta^P_g \eta^C_g  (-)^{j_g + S_{\bar q q}}
  \vert \vec R n_g; j_g |\Lambda| \xi; \eta_Y \rangle
\end{equation}

\noindent
After dividing out the quark portion of $PC$ we are left with 

\begin{equation}
(P C)_g   \vert \vec R n_g; j_g |\Lambda| \xi; \eta_Y \rangle = \xi \eta^P_g \eta^C_g  
(-)^{j_g + 1 }
  \vert \vec R n_g; j_g |\Lambda| \xi; \eta_Y \rangle
\end{equation}

\subsection{Results}

The  Salpeter equation is solved by expanding the radial wavefunction in a 
complete basis and by diagonalizing the resulting Hamiltonian matrix. The 
evaluation of the matrix elements is greatly facilitated by performing the 
angular integrals analytically. Furthermore, we found it expedient to do
the remaining numerical integrals with the potential in position space since
the integrand is less oscillatory for large argument in this case. The 
plane waves
and potential were expanded in a double series of Wigner functions yielding
a total of ten Wigner functions. The angular integral then evaluates 
to a sum over a product of six Clebsch-Gordon coefficients.

As discussed above eigenstates of the Hamiltonian may be labelled with the projection of the 
angular momentum onto the $q\bar q$ axis,  the product of gluonic parity with  
charge conjugation, $(PC)_g$  and the  Y-parity eigenvalue. 
For $\Lambda \ne 0$ the two Y-parity 
eigenstates are degenerate.  States are therefore denoted by 
$\Lambda_{(PC)_g}^Y$
where $\Lambda = 0,1,2$ are denoted by $\Sigma$, $\Pi$, $\Delta$; $(P C)_g  = 
 \eta^P_g \eta^C_g \xi (-)^{j_g+1}$ = 
$g$ or $u$ for even or
odd parity respectively; and $Y = \xi \eta^P_g \eta_Y (-)^{\Lambda+1}$ = $\pm$.
The total gluonic angular momentum, $j_g$  is not a good quantum number. 
We have found however, that the sum over radial basis states and gluon angular
momentum saturates quickly. For example, Fig 1. shows the (approximately) exact wavefunction 
for $\Sigma_g^+$ with $j_g = 1$ and a variational single Gaussian orbital.
The next contribution to the $\Sigma_g^+$ 
eigenstate has $j_g=3$, this is shown as the dotted line in the figure. 
Evidently
the eigenstate is dominated by the lowest gluonic angular 
momentum component and the single Gaussian approximation is quite accurate.
This remains true for all $R$ studied here, with the $j_g=3$ component
rising to 13\% of the wavefunction for $R = 10$ GeV$^{-1}$. These observations
support traditional approximations made for hybrids: the use of simple Gaussian
wavefunctions \cite{pss} and the truncation of the adiabatic Schr\"odinger
equation to lowest $j_g$\cite{jkm2}.

\hbox to \hsize{%
\begin{minipage}[t]{\hsize}
\begin{figure}
\epsfxsize=4in
\hbox to \hsize{\hss\epsffile{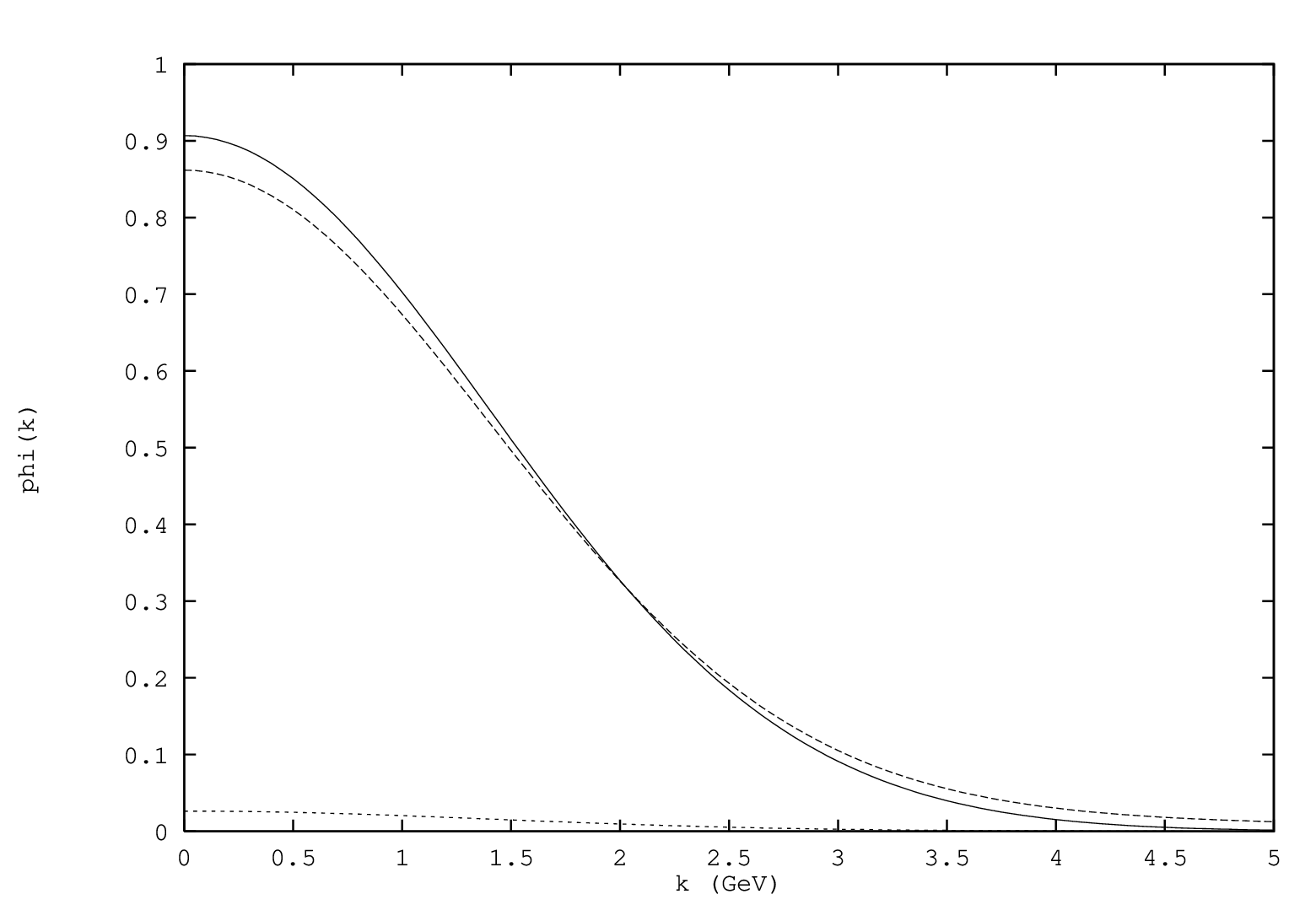}\hss}
\label{fig:1}
\end{figure}
\end{minipage}}
\begin{center}
{\small Fig.~1. $\Sigma_g^+$ wavefunctions. Exact result (dashed line), single
Gaussian approximation (solid line), $j=3$ component (dotted line).}
\end{center}

The results for the gluon spectrum are presented as a function of the 
static quark separation in Figs. 2 and 3. 
They are plotted in terms of the lattice scale, $R_0$ = 2.32 GeV$^{-1}$\cite{jkm} and the potentials have
all been normalized by subtracting an overall constant given by $V_{q\bar q}(2 R_0)$. 
In Figs.~2a-c we compare the 
recent lattice results from Ref.~\cite{jkm} 
with the predictions of the flux tube model~\cite{IP}.  
The flux tube model was motivated by the strong coupling limit of the QCD lattice Hamiltonian. 
It is based upon a nonrelativistic, small oscillation approximation to motion of the colored links  
in a topological sector where there are no overlapping (color representations of higher dimension) 
or disconnected links. Under parity (charge) conjugation the spatial (color) orientation 
of lattice links is reversed, thus the nonrelativistic ``beads'' of the flux tube model are assumed 
to flip orientation with respect to the position of the 
quarks  under parity and to have positive intrinsic charge conjugation parity.

The lowest solid line in Fig. 2a is a fit to the $q\bar q$ potential given by 
$E(R) = -4 \alpha_S/(3 R) + b R + 
const$ which corresponds to the ground state lattice $\Sigma^+_g$ potential. 
The other solid lines in Fig. 2 show the flux tube potential as given by 

\begin{equation}
E(R) = b R +  {N\pi\over R} (1 - e^{-f b^{1/2} R})
\end{equation}

\noindent
with $f \sim 1$ and 
$N=\sum_{m=1}m (n_{m+} + n_{m-})$. The latter represents the total number of right-handed ($n_{m+}$) and 
left-handed ($n_{m-}$) transverse phonon modes weighted by the phonon momentum ($m$). We note that 
the authors of Ref.~\cite{IP} included a Coulomb term in this expression, which is incompatible with the lattice results.
The flux tube model predicts the first excited $\Sigma'^+_g$ to be split by 
$N=2$ 
from the ground states, and two degenerate $\Sigma^+_u$ and $\Sigma^-_u$ potentials at $N=3$. 
In the flux tube model the lowest $\Pi$ state is predicted to be the $\Pi_u$. It is split 
from the $q \bar q$ ground state by $N=1$ and is followed by the $N=2$, $\Pi_g$ potential. 
The two lowest $\Delta$ states, $\Delta_g$ and $\Delta_u$ correspond to $N=2$ and $N=3$ respectively. 
 
The flux tube model fits the first excited state, $\Pi_u$ quite well 
over a wide range of the quark separation. It is, however, the only surface to 
do so at small distance. Furthermore,
this may be a fluke due to the particular choice of the short distance cutoff of the 
$\pi/R$ term employed in Ref.~\cite{IP}. Alternatively, at large distances, the system is expected to act
like some sort of string with an excitation energy given by $\pi/R$. The splittings
do indeed appear reasonable for all the surfaces considered. It is, however,  disconcerting that the 
$\Pi$ surfaces diverge from the flux tube model predictions for $r \gtrsim 4 R_0$. This must be
taken as an indication that the simple flux tube model considerations fail for more complex gluonic configurations.

In Fig.3a-c we plot the results of our calculations for $\Sigma$, $\Pi$ and 
the $\Delta$ potentials with the flux tube results (solid lines) for comparison.
It is apparent that the $\Pi_g$ surface lies below the $\Pi_u$ surface while
the $\Delta_u$ lies below the $\Delta_g$. This is opposite to the lattice
results, indicating that the model fails to reproduce the expected level 
orderings. The $\Sigma$ levels are also permuted with respect to the lattice
calculations at small $R$. We note, however, that the correct level 
orderings may be reinstated by simply flipping the intrinsic parity of
the gluon (set $\eta^P_g = 1$ in the expressions 
above). 
The resulting surfaces agree reasonably  well with
the lattice at small distances. For example, the $\Pi_u$ and $\Pi_g$ potentials
are roughly 1 and 2 at the origin in both calculations. We obtain values of
approximately 3 and 3.5 for the $\Delta_u$ and $\Delta_g$ surfaces at the
origin, similar to the lattice values of 2 and 3 respectively.
The agreement persists to intermediate $r$, after which it is apparent that
the model results approach the linear regime much too quickly with respect to 
the lattice. This is a strong 
indication that more degrees of freedom become active as the quark separation 
increases beyond $r \sim R_0 \sim 1/2$ fm.
This is certainly sensible from the flux tube model point of view, where a 
large number of degrees of freedom 
are necessary to construct string phonons.

Our results include the contribution due to direct interaction between the 
static quarks (the last term of Eqn. (5)). This term was set to zero by Horn and Mandula\cite{HM} 
because it anti-confines and therefore may be expected to not produce a flux tube between the
quarks. However, we note that it is responsible for producing a hybrid potential slope equal to 
that of the $q\bar q$ ground state potential, in keeping (roughly) with the lattice data. Thus we
have chosen to retain this term. Note however, that this implies that there is a short distance
repulsive Coulomb spike which should appear at very short quark separation. The appearance
of this spike is, however, unphysical because the hybrid may emit a gluon and convert into a 
$q\bar q$ color singlet and a low lying glueball and this is energetically favorable for small $r$.
Equivalently, $V_{\Pi_u} > V_{q \bar q} + m_{\rm gb}$ for $r \lesssim 0.2$ fm. Thus the Coulomb spike
mentioned above is irrelevant and one should see, instead, a Coulomb core for small $r$. This effect
may be easily incorporated into the model presented here by allowing for coupling to the $q \bar q  gg$
channel.  Notice that no core is visible in the lattice data (especially in $\Pi_u$, which is measured down to 
$r \approx 0.04$ fm).  The authors of Ref. \cite{jkm} are currently
examining this issue\cite{cj}. 

\hbox to \hsize{%
\begin{minipage}[t]{0.5\hsize}
\begin{figure}
\epsfxsize=3.5in
\hbox to \hsize{\hss\epsffile{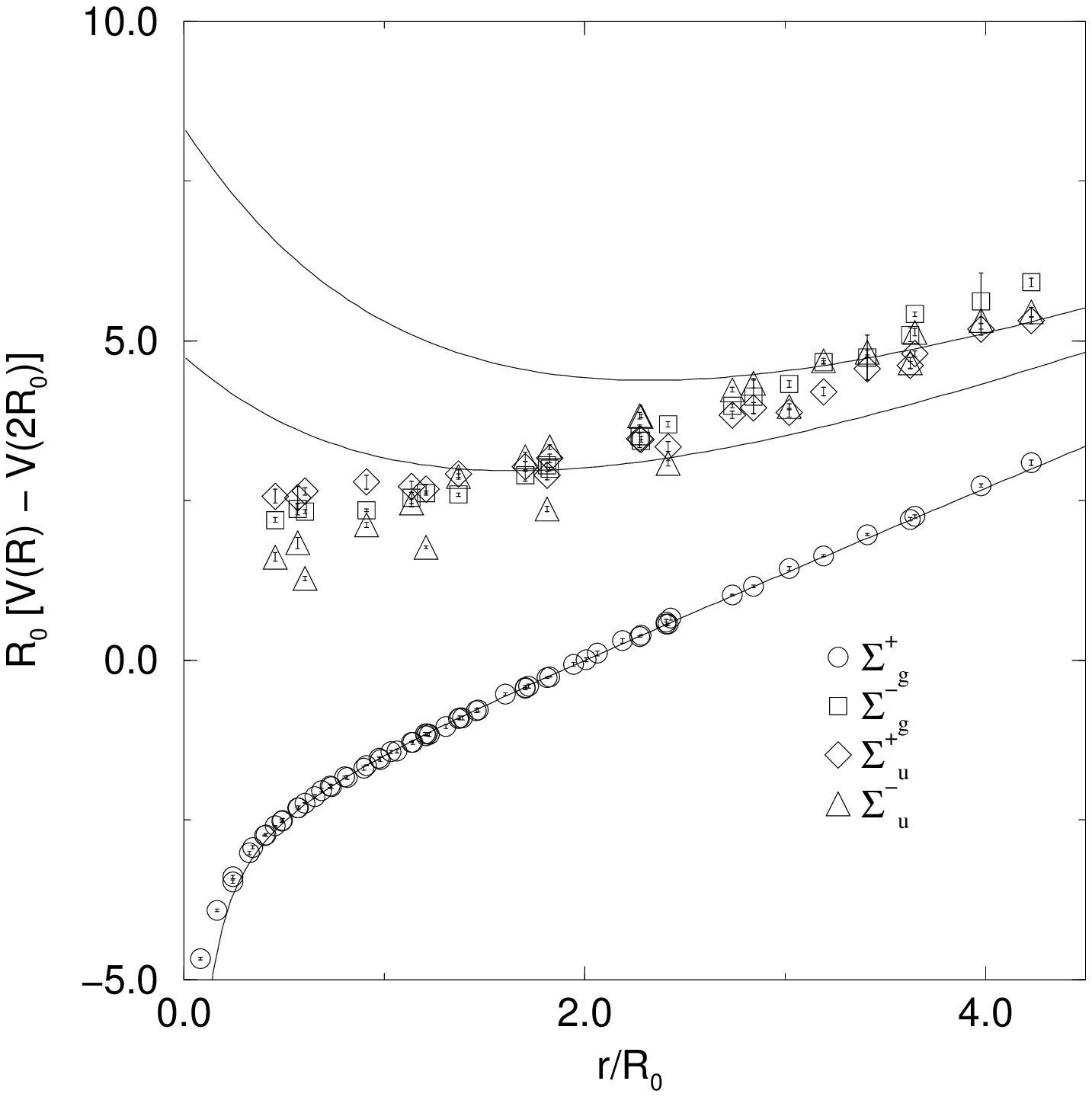}\hss}
\end{figure}
\begin{center}
{\small Fig. 2a. $\Sigma$ surfaces. Lattice (symbols) and flux tube (lines).}
\end{center}
\end{minipage}
\begin{minipage}[t]{0.5\hsize}
\begin{figure}
\epsfxsize=3.5in
\hbox to \hsize{\hss \epsffile{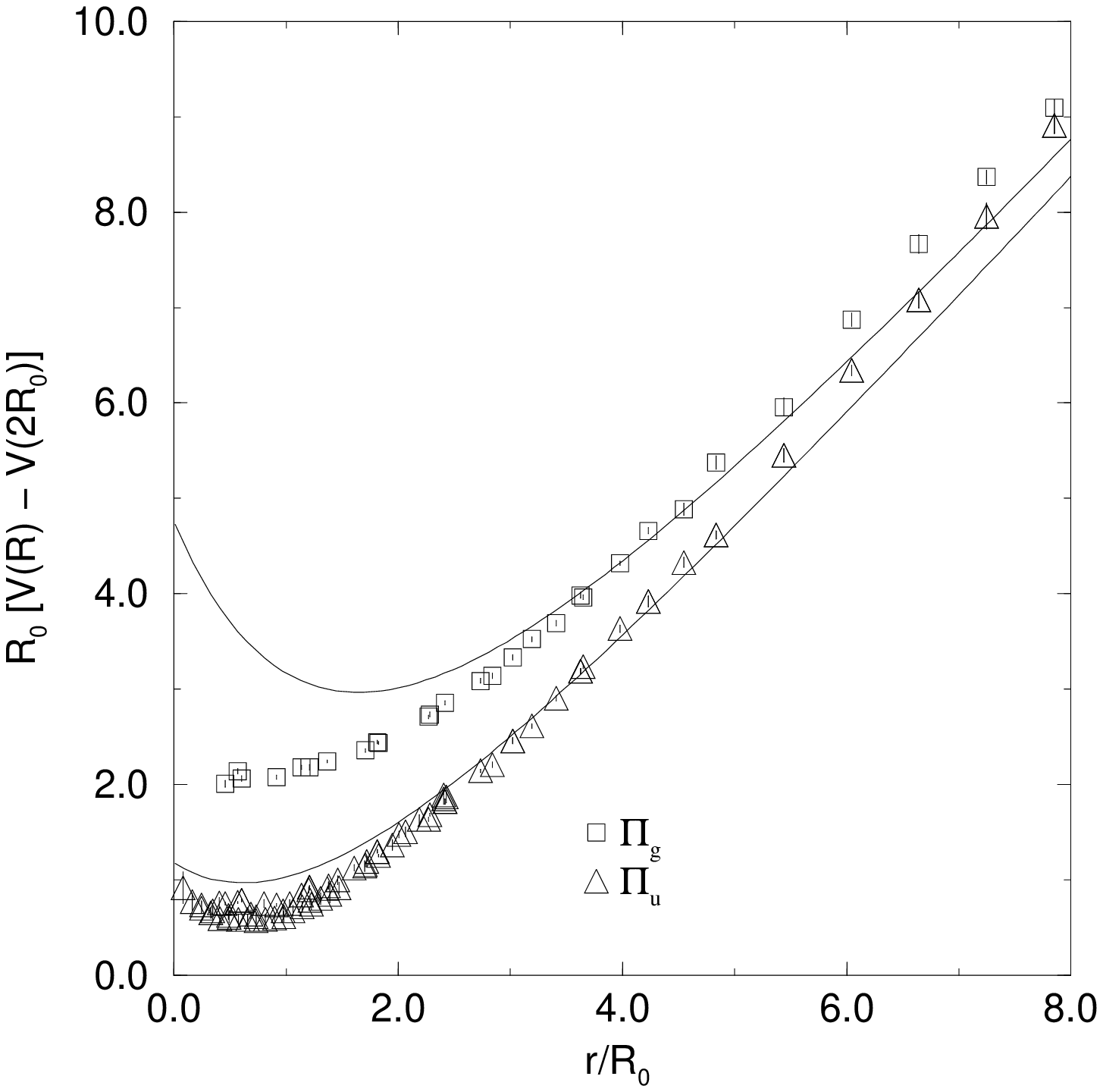}\hss}
\end{figure}
\begin{center} {\small
Fig. 2b. $\Pi$ surfaces. Lattice (symbols) and flux tube (lines). }
\end{center}
\end{minipage}}

\hbox to \hsize{%
\begin{minipage}[t]{0.5\hsize}
\begin{figure}
\epsfxsize=3.5in
\hbox to \hsize{\hss\epsffile{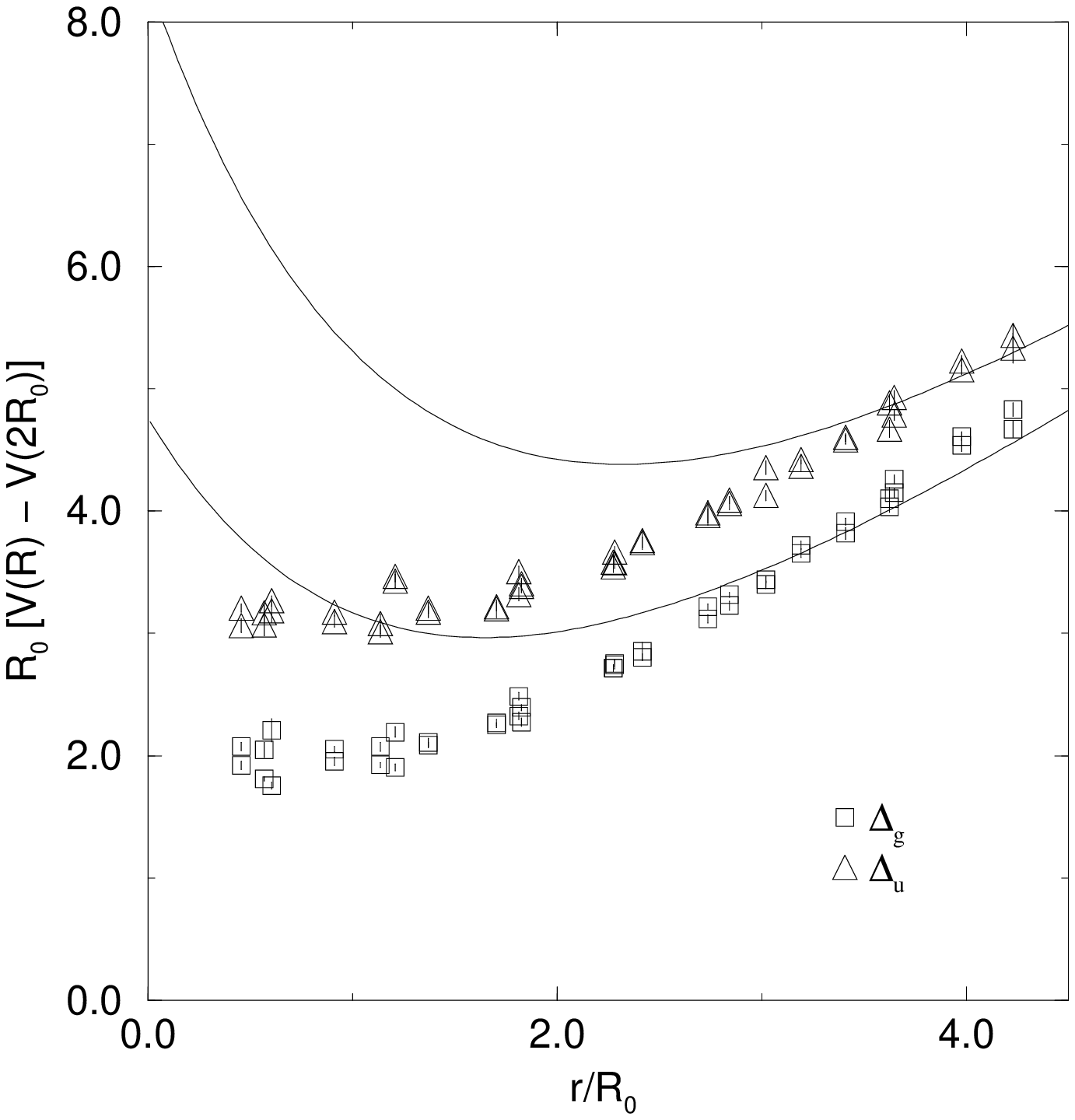}\hss}
\end{figure}
\begin{center}
{\small Fig. 2.c. $\Delta$ surfaces.  Lattice (symbols) and flux tube (lines).}
\end{center}
\end{minipage}
\begin{minipage}[t]{0.5\hsize}
\begin{figure}
\epsfxsize=3.5in
\hbox to \hsize{\hss \epsffile{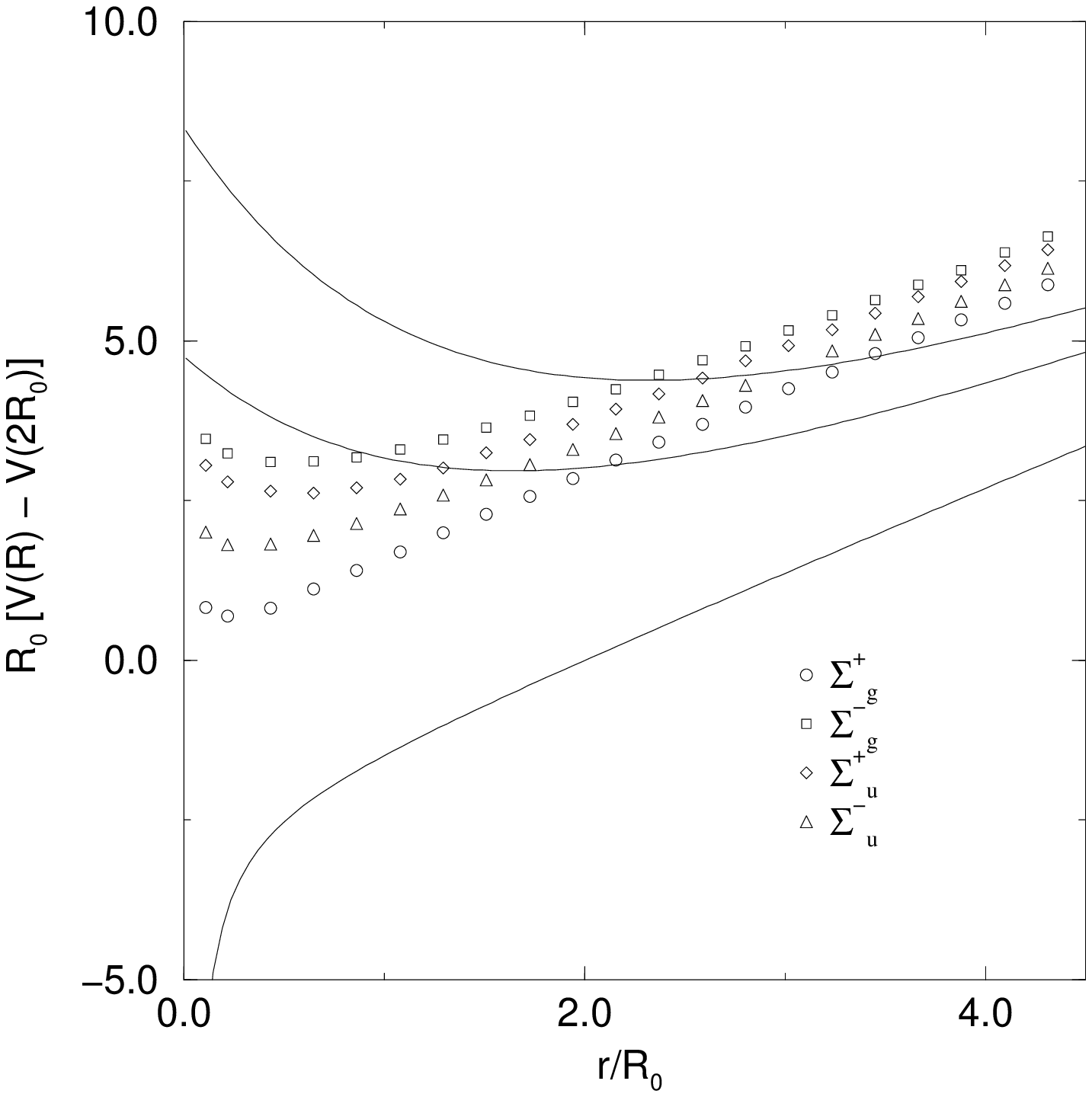}\hss}
\end{figure}
\begin{center}
{\small Fig. 3a. $\Sigma$ surfaces. Model (symbols) and flux tube (lines).}
\end{center}
\end{minipage}}

\hbox to \hsize{%
\begin{minipage}[t]{0.5\hsize}
\begin{figure}
\epsfxsize=3.5in
\hbox to \hsize{\hss\epsffile{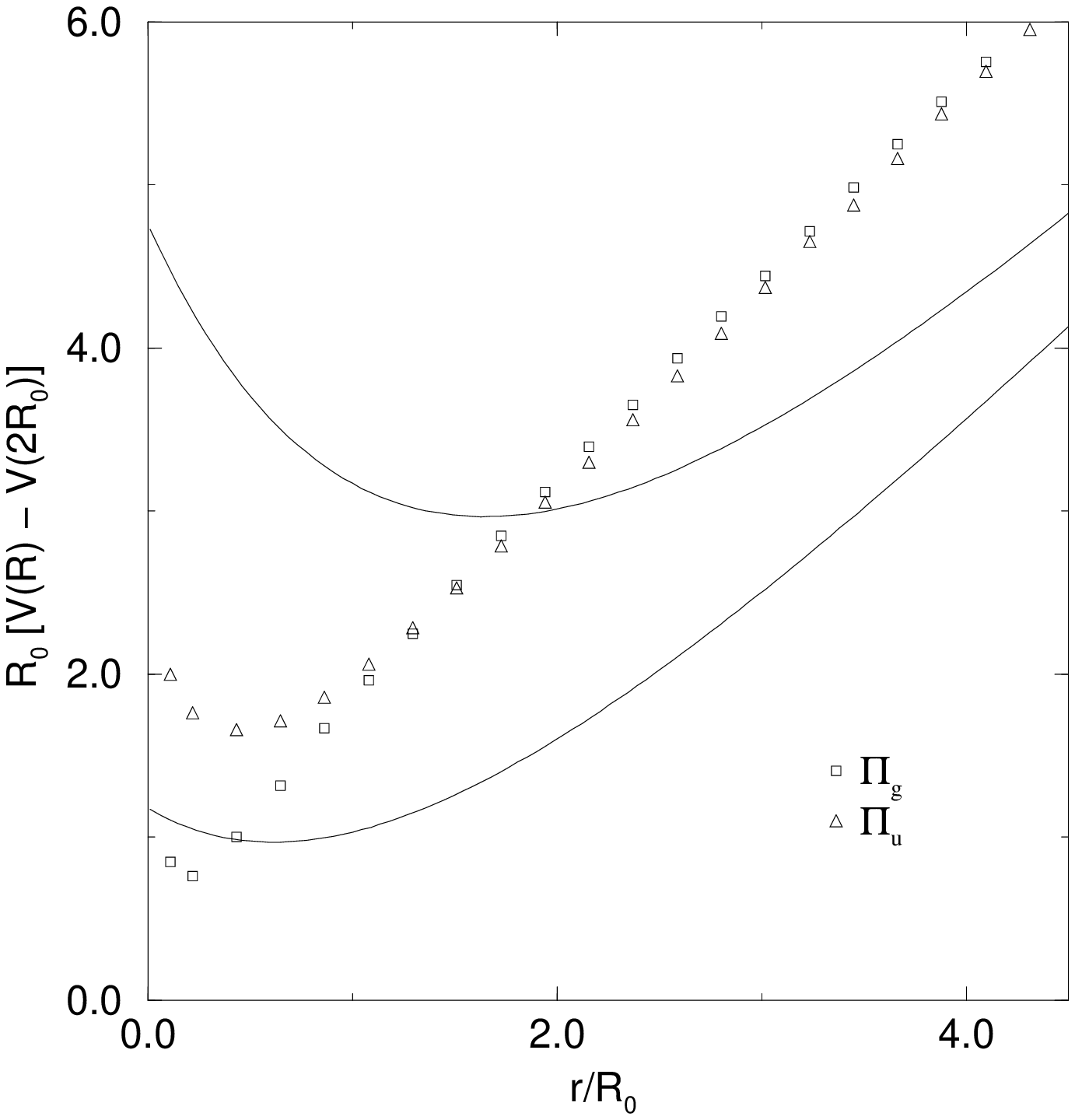}\hss}
\end{figure}
\begin{center}
{\small Fig. 3b. $\Pi$ surfaces.  Model (symbols) and flux tube (lines).}
\end{center}
\end{minipage}
\begin{minipage}[t]{0.5\hsize}
\begin{figure}
\epsfxsize=3.5in
\hbox to \hsize{\hss\epsffile{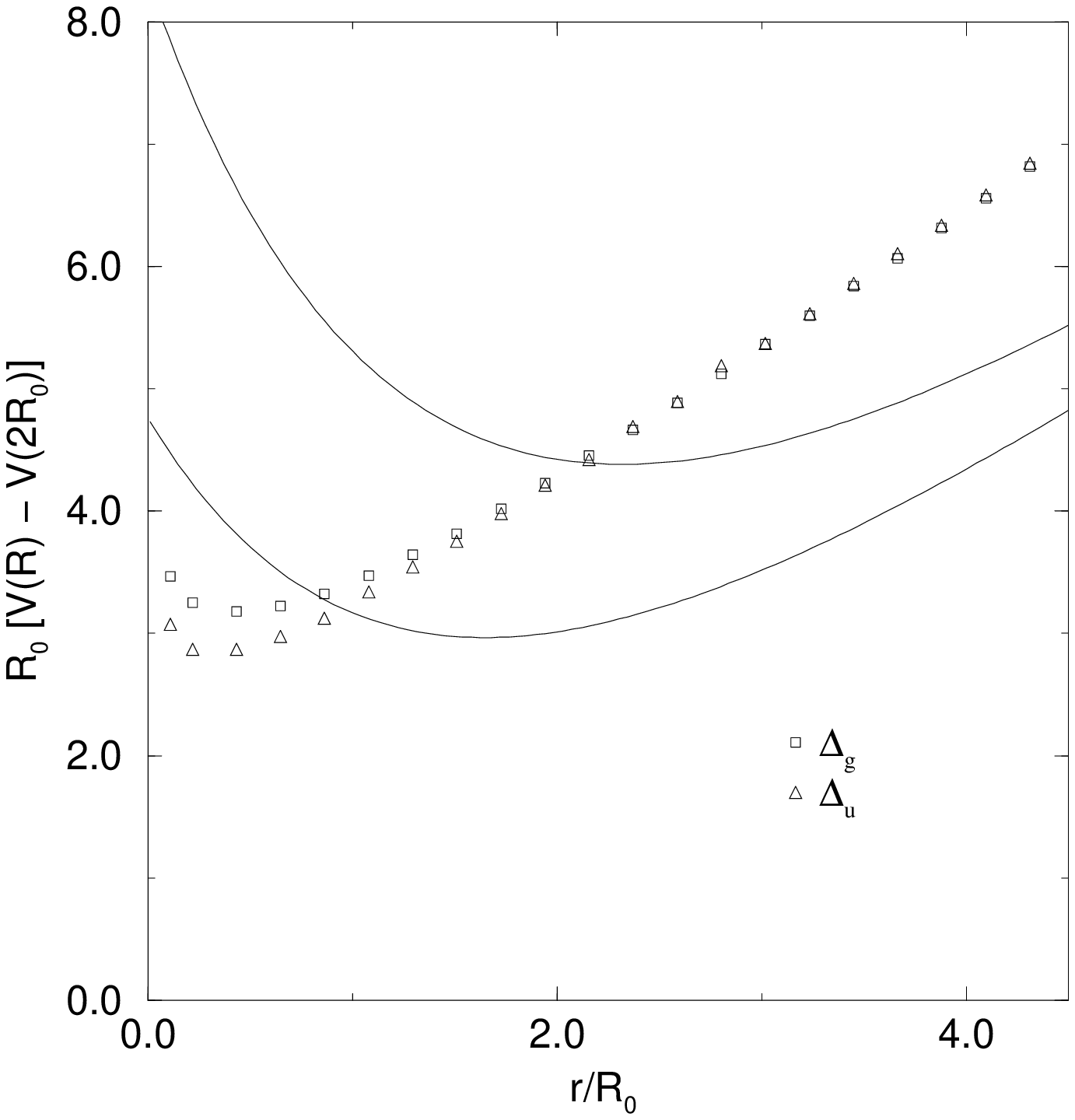}\hss}
\end{figure}
\begin{center}
{\small Fig. 3c. $\Delta$ surfaces.  Model (symbols) and flux tube (lines).}
\end{center}
\end{minipage}}

\section{Conclusions}

The model presented here agrees moderately well with lattice calculations of adiabatic
potential surfaces.  However, it disagrees in detail. In particular, it is necessary to 
ignore the intrinsic parity of the gluon to obtain the expected level
orderings. Although one may argue that this is in keeping with the expectations for the parity
of a lattice link operator (such as is employed in the development of the Flux Tube model), the
lack of consistency is disconcerting. Furthermore, the model predicts surfaces which become
linear too quickly with respect to the lattice. We expect this flaw to persist in all 
constituent glue potential models (such as \cite{HM} and \cite{bcs}).  This is a
strong indication that more degrees of freedom are required to describe soft glue at large 
interquark separation. 

It is important to note that models which employ single spinless gluons
do not contain sufficient degrees of freedom to reproduce the potential spectrum. In particular,
``single bead" Flux Tube models (\cite{bcs}) cannot make $\Pi_g$ or $\Sigma_u$ states while
three dimensional bead models (\cite{HM}) cannot make $\Sigma^+$ states. Thus including gluon
spin is a minimal necessity in this class of models (although the level ordering problem must
be overcome as discussed above).
We also note that it is possible for spherical bag models to reproduce the lattice calculations 
at small quark separation, but that they fail at large $R$. Furthermore, Flux Tube model or
Nambu-Goto string models\cite{OM} reproduce the lattice reasonably well for intermediate to 
large quark separations, but do not perform well at small $R$ or in detail at large $R$. In 
particular the potential separation does not appear to be the expected $\pi/R$ and the slopes
do not appear to agree. Finally, the bag model of Ref. \cite{HHKR,jkm2} works reasonably well 
over all quark separations, although problems remain to be resolved in the $\Sigma$ states and
for large $R$.

In summary, it appears that some sort of flux tube is required to explain the lattice 
adiabatic hybrid potential surfaces at large quark separation. This is in keeping with
the conclusions of Ref. \cite{ss3}, where flux tubes were required to explain the spin
splittings in heavy quarkonia. However, any flux tube model must attempt to incorporate 
the small $R$ behaviour of the potential surfaces (and the seemingly anomalous behaviour at
large $R$). This appears to be an indication that different degrees of freedom are required 
at small distances (as is incorporated in, for example, the bag model of Ref. \cite{HHKR}).
Finally, the model presented here (with the gluon parity reversal) works moderately well in
describing the potential surfaces and could provide a useful starting point to simple 
models of gluonic hadron properties (note that the early transition to a linear potential
seen in Figs. 3 should not be relevant since the hadron wavefunction is exponentially 
suppressed in this region).
Thus we expect that the success of the  previous glueball 
spectrum calculation\cite{ssjc} in this model (where the gluon parity was irrelevant) was
not a fluke. 
Several benefits of the model are particularly relevant to hybrids; light quarks may be easily
incorporated, the effects of coupled channels may be examined, and the effects of light quark
coupling to virtual transverse or Coulomb gluons may be included. The latter effect is of interest 
because it is excluded in quenched lattice calculations and may have a substantial effect on the
hidden flavor hybrid spectrum.

\acknowledgements 
The authors are grateful to C. Morningstar for discussions and for providing us the preliminary
lattice data of Ref. \cite{jkm}. This work was supported by the DOE 
 under grants DE-FG02-96ER40944 (ES) and DE-FG02-87ER40365 (AS).

\end{document}